\documentclass[12pt,epsf]{article}
\usepackage{amsmath}
\usepackage{amssymb,color}
\newcommand{\no}{\noindent}

\begin{document}
\def\a{\alpha}
\def\b{\beta}
\def\c{\varepsilon}
\def\d{\delta}
\def\e{\epsilon}
\def\f{\phi}
\def\g{\gamma}
\def\h{\theta}
\def\k{\kappa}
\def\l{\lambda}
\def\m{\mu}
\def\n{\nu}
\def\p{\psi}
\def\q{\partial}
\def\r{\rho}
\def\s{\sigma}
\def\t{\tau}
\def\u{\upsilon}
\def\v{\varphi}
\def\w{\omega}
\def\x{\xi}
\def\y{\eta}
\def\z{\zeta}
\def\D{{\mit \Delta}}
\def\G{\Gamma}
\def\H{\Theta}
\def\L{\Lambda}
\def\F{\Phi}
\def\P{\Psi}

\def\S{\Sigma}

\def\o{\over}
\def\beq{\begin{eqnarray}}
\def\eeq{\end{eqnarray}}
\newcommand{\gsim}{ \mathop{}_{\textstyle \sim}^{\textstyle >} }
\newcommand{\lsim}{ \mathop{}_{\textstyle \sim}^{\textstyle <} }
\newcommand{\vev}[1]{ \left\langle {#1} \right\rangle }
\newcommand{\bra}[1]{ \langle {#1} | }
\newcommand{\ket}[1]{ | {#1} \rangle }
\newcommand{\EV}{ {\rm eV} }
\newcommand{\KEV}{ {\rm keV} }
\newcommand{\MEV}{ {\rm MeV} }
\newcommand{\GEV}{ {\rm GeV} }
\newcommand{\TEV}{ {\rm TeV} }
\newcommand\red[1]{{\textcolor{red}{#1}}}
\def\slash#1{\ooalign{\hfil/\hfil\crcr$#1$}}
\def\diag{\mathop{\rm diag}\nolimits}
\def\Spin{\mathop{\rm Spin}}
\def\SO{\mathop{\rm SO}}
\def\O{\mathop{\rm O}}
\def\SU{\mathop{\rm SU}}
\def\U{\mathop{\rm U}}
\def\Sp{\mathop{\rm Sp}}
\def\SL{\mathop{\rm SL}}
\def\tr{\mathop{\rm tr}}

\baselineskip 0.7cm

\begin{titlepage}

\begin{flushright}
UCB-PTH-09/10
\end{flushright}

\vskip 1.35cm
\begin{center}
{\large \bf Unparticles and Holographic Renormalization Group}

\vskip 2.2cm

Chiu Man Ho and Yu Nakayama

cmho@berkeley.edu

nakayama@berkeley.edu

Department of Physics, University of California, Berkeley, CA 94720

Theoretical Physics Group, Lawrence Berkeley National Laboratory,
Berkeley, CA 94720

\vskip 2.5cm

\abstract{ We revisit the unparticle interactions and propagators
from the AdS-CFT point of view, and we show how the contact terms
and their renormalization group flow appear in the context of the
holographic renormalization. We study both vector unparticles and
unfermions, uncovering the relevant boundary conditions and
renormalization group flows. }

\end{center}
\end{titlepage}

\setcounter{page}{2}

\section{Introduction}

Unparticles have very peculiar properties compared with ordinary
particles. In his pioneering work, Georgi \cite{Georgi:2007ek}
defined unparticles as the ``(approximately) scale invariant field
theory that weakly couples with the standard model sector". The most
important properties of unparticles is its scale
invariance.
The scale invariance might be imposed around the electro-weak energy
scale, where we hope to find new physics in near-future experiments.
The approximate scale invariance means that the scale invariance
might (or might not) be broken at much higher (or lower) energy
scale than the energy scale $E$ that we would like to observe the
unparticle.

As a simple example of the unparticle sector, Georgi considered
Bank-Zaks (BZ) type conformal field theory (CFT)
\cite{Banks:1981nn}, which is defined by QCD with many massless
fundamental fermions. Below the dynamical scale
$\Lambda_{\mathcal{U}}$ of QCD, the theory is approximately
conformal. If we introduce masses for the fermions, the conformal invariance would be
broken at energy scale lower than $\Lambda_{{\scriptsize{\slash
{\mathcal{U}}}}}$. The approximate scale invariance demands the
inequality $ \Lambda_{{\scriptsize {\slash {\mathcal{U}} }}}\ll E
\ll \Lambda_{\mathcal{U}}$. Another important scale in unparticle
physics is the mass scale $M_{\mathcal{U}}$ of the messenger fields,
at which an unparticle operator $O_{UV}$ at ultraviolet (UV) couples
with a standard model (SM) operator $O_{SM}$ as $\frac{O_{SM}
O_{UV}}{M_{\mathcal{U}}^k}$. Below the conformal scale
$\Lambda_{\mathcal{U}}$, it becomes the effective coupling between
the scale invariant field theory and the SM sector as  $
\frac{C_{\mathcal{U}}\,\Lambda_{\mathcal{U}}^{d_{UV}-d_\mathcal{U}}}{M_{\mathcal{U}}^k}
O_{SM} O_{\mathcal{U}}$, where $k=d_{UV}+d_{\mathcal{U}}-4$ with
$d_{UV}$ and $d_{\mathcal{U}}$ being the scaling dimension of
unparticle operator at UV and scale invariant fixed point
respectively.

Notice that the scaling dimension of unparticle operators is very
important because when $d_{\mathcal{U}}$ is large, the interaction
may be too weak to be observed in nature. However, if we assume the
conformal invariance in the unparticle sector, there is a severe
unitarity bound for the scaling (\,=\,conformal\,) dimension of
primary operators \cite{Mack:1975je}:

\begin{eqnarray}
 d \ge j_1 + j_2 + 2 - \delta_{j_1j_2,0}  \ ,
\end{eqnarray}

\noindent where $j_1$ and $j_2$ are Lorentz spin of the operator. As
first pointed out in \cite{Nakayama:2007qu} (see also
\cite{Grinstein:2008qk}), this unitarity bound is neglected by many
authors in the study of vector unparticles, including Georgi's
original work.

In fact, the unparticle interaction
$\frac{O_{SM}O_{UV}}{M_{\mathcal{U}}^k}$ might not be the dominant
contribution to the standard model process in new physics. For
instance, the contact term interaction
$\frac{O_{SM}^2}{M_{\mathcal{U}}^{k'}}$ introduced at the same UV
scale $M_{\mathcal{U}}$ could be the dominant piece. Indeed, in
\cite{Grinstein:2008qk}, it has been shown that such an interaction
should result from the renormalization group (RG) flow of the unparticle
operators. Denoting the new interaction as $\sqrt{B_1}
\frac{O_{SM}O_{\mathcal{U}}}{M_{\mathcal{U}}^k} + B_2
\frac{O_{SM}^2}{M_{\mathcal{U}}^{k'}}$, they have shown that the
Callan-Symanzik equation gives

\begin{eqnarray}\label{CSE}
\left(\frac{\partial}{\partial \log \mu} + \beta(g)
\frac{\partial}{\partial g}\right) B_i = \gamma_{ij}(g) \,B_j \ ,
\end{eqnarray}
where $\gamma_{ij}$ are the anomalous dimension matrix. The solution
to the RG equation can be obtained as

\begin{eqnarray} \label{B_one}
B_1(\mu)&=&\left(\frac{\mu}{\Lambda_{\mathcal{U}}}\right)^{\gamma_{11}(g_{\ast})}
B_1(\Lambda_{\mathcal{U}})\\ B_2(\mu) &=& B_2(\Lambda_{\mathcal{U}})
+ \frac{\gamma_{12}(g_*)}{\gamma_{11}(g_*)}
\left[\left(\frac{\mu}{\Lambda_{\mathcal{U}}}\right)^{\gamma_{11}(g_*)}
- 1\right]\, B_1(\Lambda_{\mathcal{U}}) \,, \label{B_two}
\end{eqnarray}

\no where $g_{\ast}$ is the non-trivial IR fixed point associated
with the conformal sector.

Note that as discussed in \cite{Grinstein:2008qk}, the ratio between
the contribution from the unparticle exchange and the contact term
can be computed as

\begin{eqnarray}
\frac{A_{\textrm{unparticle}}}{A_{\textrm{contact}}} =
\frac{B_2^2}{\sqrt{B_1}}
\left(\frac{E}{M_{\mathcal{U}}}\right)^2\left(\frac{E}{\Lambda_{\mathcal{U}}}\right)^{2(d-3)}
\ .
\end{eqnarray}

\noindent Obviously, for $E<\Lambda_{\mathcal{U}}<M_{\mathcal{U}}$
and vector unparticles with scaling dimension $d_V \geq 3$ (as
required by unitarity), the unparticle exchange is naturally
suppressed.

In this article, we will reproduce (\ref{B_two}) by AdS-CFT
correspondence. We will also show how the contact terms and their
 RG flow appear in the context of the holographic
renormalization group. This requires a careful treatment of the
boundary terms, which is sometimes neglected in the string theory
literatures. As we will see, the boundary terms in the AdS-CFT
generate the contact term interaction in the CFT and eventually lead
to the effective standard model coupling  $B_2
\frac{O_{SM}^2}{M_{\mathcal{U}}^{k'}}$. The holographic
renormalization group equation will give the counterpart of
\eqref{B_two} in the CFT sector.

\section{Vector Unparticles Revisited}

One interesting theoretical approach to unparticle physics is to use
AdS-CFT correspondence
\cite{Stephanov:2007ry,Lee:2007xd,Strassler:2008bv,Cacciapaglia:2008ns,Friedland:2009iy}.\footnote{In
this section, we would like to assume conformal invariance rather
than mere scale invariance. The geometric description with only
scale invariance is an interesting direction but it is not well
understood. Maybe there is a geometrical way to prove or disprove
the equivalence between conformal invariance and scale invariance in
higher dimension.} The basic statement of the AdS-CFT correspondence
is that a strongly coupled conformal field theory can be analysed by
a weakly coupled gravitational theory on AdS space. Although there
is no known way to represent the gravity dual for SQCD (or
Banks-Zaks theory), many other non-trivial superconformal field theories can be analysed
from gravity.

It is rather trivial to see that both theories possess the same
symmetry: on the CFT side, we have conformal $SO(2,4)$ symmetry
while the AdS space has an isometry group given by $SO(2,4)$. In
particular, under this correspondence, the AdS global energy
(Hamiltonian) corresponds to the conformal dimension of CFT
operators. In the following, we mainly consider the AdS space in the
Poincare coordinate
\begin{eqnarray}
ds^2_{AdS} = \frac{dz^2 + dx^{\mu} dx_{\mu}}{z^2} \ ,
\end{eqnarray}

\noindent where the radial direction $z$ corresponds to the energy
scale of the CFT.

In addition to this kinematical correspondence, AdS-CFT  predicts a
dynamical relation  (known as GKPW relation \cite{Gubser:1998bc,
Witten:1998qj}) between the generating functions of the CFT
correlation functions and the path integral for the gravitational
theory with fixed boundary condition:

\begin{eqnarray}\label{ADS_CFT}
Z_{AdS}[A_{0,\mu}] = \int_{A_{M}|_{boud}=A_{0,\mu}} \mathcal{D}A_M
\exp(-I[A_M]) \equiv Z_{CFT}[A_{0,\mu}] = \left\langle \exp (\int
d^4 x A_{0,\mu}O^{\mu} )\right\rangle  \ ,
\end{eqnarray}

\noindent where $A_{0,\mu}$ is a suitably defined boundary value of
the 5-dimensional vector field $A_{M}$ and $O^{\mu}$ is the
corresponding source current in the CFT.  Later, we will use this
relation to compute the unparticle propagator.

The unparticle hidden sector is not an idealistic CFT, however. At
least we need (non-conformal) coupling between the hidden sector and
the SM sector. We may also want to introduce IR cut-off (or relevant
deformation) below the electro-weak scale. In the AdS-CFT language,
this field theory cut-off can be understood as a modification of the
geometry at UV (or IR). We can introduce UV brane at $z= z_{UV}
=\frac{1}{M_{\mathcal{U}}}=\epsilon$ to mimic the coupling to the SM
sector. The IR cut-off can be also introduced by capping off the
geometry at $z=
z_{IR}=\frac{1}{\Lambda_{{\scriptsize{\slash{\mathcal{U}} }}}}$. The
construction is much like the Randall-Sundrum scenario (see e.g.
\cite{Perez})\,; and it is known as ``unparticle deconstruction"
\cite{Stephanov:2007ry}.

In this section, we will revisit the vector unparticle propagator
from the AdS-CFT point of view by focusing on the contact term
interactions whose importance was emphasized in
\cite{Grinstein:2008qk}. The contact terms are neglected in most
applications of AdS-CFT because one can remove them by local counter
terms in the boundary action. They are important, however, because
they will affect the unparticle physics by introducing effective
higher dimensional operators such as

\begin{eqnarray}
L_{\textrm{eff}} = C_0 j_{\mu} j^{\mu} + C_1 j_{\mu} \partial^2
j^{\mu} + C_2 (\partial^{\mu} j_{\mu})^2 + \cdots \label{inn}
\end{eqnarray}

\noindent in the standard model Lagrangian ($j_\mu$ is a current in
the standard model).

The correspondence between the contact terms in the CFT and the
higher dimensional interactions in the standard model Lagrangian can
be understood as follows. We assumed the interaction $\frac{O_{SM}
O_{UV}}{M_{\mathcal{U}}^k}$, so if we have a contact term
interaction between $O_{\mathrm{\mathcal{U}}}$, then the
perturbative expansion of the standard model-unparticle interaction
generates $ \int d^4y \, O_{SM}(y) O_{SM}(x)\,\langle
O_\mathcal{U}(y) O_{\mathcal{U}}(x) \rangle$, which will yield us a
higher derivative interaction $B_2
\frac{O_{SM}^2}{M_{\mathcal{U}}^{k'}}$ in \eqref{inn} from the
contact terms such as $ C_0 \delta (x) + C_1
\partial^2 \delta (x) + \cdots$ (in addition to the conventional
standard model-unparticle interaction $\sqrt{B_1}
\frac{O_{SM}O_{\mathcal{U}}}{M_{\mathcal{U}}^k}$). With this
correspondence, we can identify the coupling constants $C_i$ in
\eqref{inn} as the coefficients appearing in the contact term of the
correlation functions in the CFT. We will show that the natural
 RG flow generates such terms completely in
agreement with the field theory discussion \cite{Grinstein:2008qk}.
From the higher dimensional brane scenario perspective, our
prescription provides a natural way to understand the evolution of
the boundary local counter terms under the RG flow.

First of all, the action for the 5-dimensional massive vector (5d
Proca action) is given by \cite{Mueck:1998iz}

\begin{eqnarray}
I = \int d^5x \sqrt{g}\left( \frac{1}{4}F_{M N}F^{M N} +
\frac{1}{2}{m^2}A_M A^M \ \right),
\end{eqnarray}
where $F_{M N} = \partial_{M}A_{N} - \partial_{N}A_{M}$. This leads
to the equation of motion (Proca equation)
\begin{eqnarray}
\nabla_M F^{MN} - m^2 A^N = 0\,.
\end{eqnarray}

\no By taking the divergence of the Proca equation\footnote{When
$m^2=0$, this is nothing but a Lorenz gauge condition, but for
$m^2\neq0$ it follows from the equation of motion.}, we obtain the
divergence free condition
\begin{eqnarray}
\nabla_M A^M = 0\,.
\end{eqnarray}

This action can be evaluated by the boundary data
$\tilde{A}_{\epsilon,\mu}(k)$, which is the Fourier transform of the
Dirichlet boundary value of the field $\tilde{A}_{\mu}(k)$ at
$z=\epsilon$:

\begin{eqnarray}
I &=& \epsilon^{-4}\,\frac{d-3}{2}\,\int \frac{d^4 k}{(2\pi)^4}
\,\tilde{A}_{\epsilon,\mu} \tilde{A}_{\epsilon,\mu} \nonumber \\
&&-\frac{\epsilon^{-2}}{4}\, \frac{\Gamma(d-3)}{\Gamma(d-2)} \int
\frac{d^4 k}{(2\pi)^4} \,k^2\,
\tilde{A}_{\epsilon,\mu}\left(-\delta_{\mu\nu} +
\frac{2(d-2)}{d-1}\frac{k_\mu k_\nu}{k^2}
\right)\tilde{A}_{\epsilon,\nu} \nonumber \\ &&-
\frac{\epsilon^{2(d-4)}}{4^{d-2}}\, \frac{\Gamma(3-d)}{\Gamma(d-2)}
\int \frac{d^4k}{(2\pi)^4}\,(k^2)^{d-2}
\,\tilde{A}_{\epsilon,\mu}\left(-\delta_{\mu\nu} +
\frac{2(d-2)}{d-1}\frac{k_\mu k_\nu}{k^2}
\right)\tilde{A}_{\epsilon,\nu} + \cdots, \label{contt}
\end{eqnarray}

\no where higher derivative terms with higher order $\epsilon$ is
neglected. For later purposes, however, we have incorporated the
contact terms  neglected in \cite{Mueck:1998iz}.\footnote{The
conformal invariance does not fix the structure of the contact
terms, so this is the reason why they are often neglected in the
literatures of AdS-CFT correspondence. Here, we show that the
boundary counter terms play an important role to determine the
contact terms and their RG-flow.}

The mass $m$ in the 5d-bulk space is related to the conformal
dimension $d$ of the dual operator as $d =2+ \sqrt{1+m^2}$
\cite{Mueck:1998iz}. Generalizing the discussion in
\cite{Witten:1998qj}, one can easily see that for the vector
particle, the stability bound is $m^2 \geq 0$, corresponding to the
unitarity bound for the vector unparticle $d_V \geq 3$. The
necessity of the unitarity bound can also be seen as the requirement
of the (Euclidean) non-normalizabity of the wave under the inner
product $\int d^5x \sqrt{g}\, g^{MN} A_{M}A_{N}$ with $A_\mu(z) \sim
z^{4-d}$ near $z\sim 0$.

The third line in \eqref{contt}, which is in general non-analytic,
will reproduce the CFT two-point function \cite{Grinstein:2008qk}
(up to a normalization factor $c$)

\begin{align}
\langle O_\mu(x) O_\nu(0) \rangle &= \frac{c}{2\pi^2}
\frac{\delta_{\mu\nu} - 2x_\mu x_\nu/x^2}{(x^2)^d} \cr &= c\,
\frac{(d-1)\Gamma(2-d)}{4^{d-1}\Gamma(d+1)} \int \frac{d^4
k}{(2\pi)^4}\, e^{ikx}(k^2)^{d-2}\left(\delta_{\mu \nu}-
\frac{2(d-2)}{d-1}\frac{k_\mu k_\nu}{k^2}\right) \label{twop}
\end{align}

\no from the AdS-CFT prescription \cite{Gubser:1998bc,
Witten:1998qj} as shown in (\ref{ADS_CFT}) together with a suitable
analytic continuation in the Fourier integral. This is achieved by
specifying the boundary data $\tilde{A}_{0,\mu} = \lim_{\epsilon \to
0} \tilde{\mathcal{A}}_{\epsilon,\mu}$ with the normalized field $
\tilde{\mathcal{A}}_{\epsilon,\mu} \equiv \epsilon^{d-4}
\tilde{A}_{\epsilon,\mu}$.

In contrast, the first line and the second line in \eqref{contt} are
not dictated by the conformal invariance but they give contact
terms. At a given $\epsilon$, one can always eliminate such contact
terms by adding the boundary counter terms as

\begin{align}
\delta S_{bound} &= \epsilon^{-2(4-d)} \int \frac{d^4k}{(2\pi)^4}
\,\left(\,c_0\tilde{A}_{\epsilon,\mu} \tilde{A}_{\epsilon,\mu} + k^2
\,\tilde{A}_{\epsilon,\mu}\left( c_1 \delta_{\mu\nu} + c_2
\frac{k_\mu k_\nu} {k^2}\right) \tilde{A}_{\epsilon,\nu} \ + \cdots
\,\right) \cr
 &= \int \frac{d^4k}{(2\pi)^4}\,\left(\, c_0\tilde{\mathcal{A}}_{\epsilon,\mu}
 \tilde{\mathcal{A}}_{\epsilon,\mu} + k^2\,\tilde{\mathcal{A}}_{\epsilon,\mu}\left( c_1
 \delta_{\mu\nu} + c_2\frac{k_\mu k_\nu}{k^2}\right) \tilde{\mathcal{A}}_{\epsilon,\nu} + \cdots\,\right),
\end{align}

\no which are localized on the UV-brane. However, what we would like
to study here is the RG flow of the contact
terms. In other words, we would like to investigate the cut-off
dependence of the contact terms in the AdS-CFT setup.

We find that it is natural to introduce the cut-off dependence on
the boundary counter term by parameterizing $ c_0 = \tilde{C}_0
\,\epsilon^{4-2\Delta_0}$ and $ c_{1,2} = \tilde{C}_{1,2}
\,\epsilon^{6-2\Delta_0}$ , where we have introduced the ``naive
dimension" $\Delta_0$ of the current operator under consideration.
The point is that the RG equation in \cite{Grinstein:2008qk}
involves the ``anomalous" dimension which is only defined by
comparing the actual dimension of certain operators with a reference
value (say, a UV free theory). In fact, they made an assumption that
they normalize their operators with respect to the UV free theory.
Correspodingly, the cut-off dependence introduced here is normalized
so that for the free field current interaction (i.e. $\Delta_0 =3$),
$c_{1,2}$ are cut-off independent and dimensionless. Once we have
determined to evaluate the anomalous dimension of CFT operators with
respect to the free field theory by utilizing the same convention
used in \cite{Grinstein:2008qk}, the vanishing cut-off dependence
for $\Delta_0 =3$ is fixed by definition. Simple dimensional
analysis also determines other cut-off dependence like $c_0$. Now,
with different cut-offs, we have the relation:

\begin{eqnarray}\label{C_zero}
C_0(\epsilon) &=& \tilde{C}_0\,\epsilon^{4-2\Delta_0} \left(\;1-
\left(\frac{\tilde{\epsilon}_0}{\epsilon}\right)^{\gamma}\;\right)\,, \\
\label{C_one} C_1(\epsilon) &=& \tilde{C}_1\,\epsilon^{6-2\Delta_0}
\left(\; 1-
\left(\frac{\tilde{\epsilon}_1}{\epsilon}\right)^{\gamma}\;\right)\,, \\
\label{C_two} C_2(\epsilon) &=& \tilde{C}_2\,\epsilon^{6-2\Delta_0}
\left(\; 1-
\left(\frac{\tilde{\epsilon}_2}{\epsilon}\right)^{\gamma}\;\right)\,,
\end{eqnarray}

\noindent where we have introduced the anomalous dimension $\gamma =
2(d-\Delta_0)$. $\tilde{\epsilon}_i$ denotes the scale at which the
boundary counter terms cancel the bulk contributions, and they can
be different for different $i$ in principle. It is easy to see that
$C_1$ in (\ref{C_one}) is equivalent to $B_2$ in (\ref{B_two}) by
the identifications $\Delta_0=3$ and $\epsilon \sim \frac{1}{\mu}$.
In this way, we have shown how AdS-CFT correspondence also predicts
the appearance of the contact terms and their evolution.

Several comments are in order:
\begin{itemize}

\item The choice of naive dimension $\Delta_0 =3$ is natural because $\tilde{A}_{\epsilon,\mu}$ couples to the vector operator $O^\mu$, and the typical (actually the lowest dimensional) free field vector operator has dimension $3$ such as $\phi^\dagger \partial_\mu \phi$ or $\bar{\psi}\gamma_\mu \psi$.

\item Unlike the claim $(C_0=C_1)$ in \cite{Grinstein:2008qk}, $C_0$ and $C_1$ are not a-priori related though we could always relate them as a boundary condition at the cut-off. This difference is due to the fact that they implicitly assumed  the simplest weakly coupled messengers that propagate between the unparticle sector and the standard model sector. For a more general strongly coupled mediation, the condition $(C_0=C_1)$ will be generically violated.

\item When $d$ is an integer, the distinction between the boundary counter terms (contact terms) and the bulk term is less clear because the propagator is analytic in $k$. This is somehow related to the artificial divergence of some unparticle amplitudes at integer value of $d$ appearing in the literatures \cite{Grinstein:2008qk}. It simply suggests that the normalization of the operator is not good: we can always remove the divergence by the counter term or proper choice of the renormalized coupling constant.

\item Since the AdS gravity dual does not know anything about the ``anomalous" dimension but only knows the ``actual" conformal dimensions, the introduction of the naive dimension as a regularization (boundary counter term) is necessary. Our prescription is the most natural one in the sense that it is in complete agreement with the field theory. In principle, we could embed the whole system inside an ``asymptotically free field dual" of the gravity theory to discuss the anomalous dimensions and operator evolution without using somewhat artificial boundary counter terms. Our prescription, however, should be an effective way to implement this hypothetical procedure because there is no known simple gravity dual for  asymptotically free field theories.

\end{itemize}

\section{Unfermions}

A similar construction is also possible for the Dirac field
(unfermion) \cite{Cacciapaglia:2008ns}. The 5d action is given by
\cite{Mueck:1998iz}

\begin{eqnarray}
\int d^5 x \sqrt{g}\, \bar{\psi}(\slash{D}-m)\psi + G\int d^4
x\sqrt{h}\, \bar{\psi}\psi \,,
\end{eqnarray}

\no which is supplemented with a surface term \cite{surface-term}
with an undetermined coefficient $G$.

The bulk Dirac equation
\begin{eqnarray}
(\slash{D}-m)\psi = 0
\end{eqnarray}

\no gives the relation between the left-mover $\psi^+_\epsilon$ and
right-mover $\psi^-_\epsilon$ at the boundary $z=\epsilon$, and the
boundary action can be determined solely from the boundary term. The
action is

\begin{eqnarray}\label{I_unfermion}
I = i\,G\,\epsilon^{-3}\;\frac{\Gamma(d_F-\frac52)}{\Gamma
(d_F-\frac32)}\, \int\frac{d^4k}{(2\pi)^4}
\,\left(\,\bar{\psi}^{+}_\epsilon k_\mu \gamma^\mu \psi^{-}_\epsilon
- \frac{\Gamma(d_F-\frac12)}{\Gamma
(d_F-\frac32)}\,\left(\frac{k\,\epsilon}{2}\right)^{2\,d_F-5}
\bar{\psi}^{+}_\epsilon k_\mu\gamma^\mu \psi^{-}_\epsilon  + \cdots
\right),
\end{eqnarray}
where $d_F=m+2$ is the scaling dimension of the unfermions
\cite{Mueck:1998iz}.\footnote{We concentrate on $m \geq 1/2$ here
for simplicity. The case $m \leq -1/2$ can be treated similarly but
with the roles of $\psi^+_\epsilon$ and $\psi^-_\epsilon$ exchanged
\cite{Mueck:1998iz}. See also \cite{Cacciapaglia:2008ns} for
$|m|<1/2$.}

The second term in (\ref{I_unfermion}) will generate the correct
unfermion propagator as follows. Let $\chi^+$ and $\bar{\chi}^-$ be
the boundary spinors which couple to $\bar{\psi}^+_0$ and $\psi_0^-$
respectively, where

\begin{equation}
\bar{\psi}^+_0=\lim_{\epsilon \rightarrow 0} \,\epsilon^{\,d_F-4}\,
\bar{\psi}^{+}_\epsilon ~~~~\textrm{and} ~~~~\psi_0^- =
\lim_{\epsilon \rightarrow 0} \,\epsilon^{\,d_F-4}\,
\psi^{-}_\epsilon\,.
\end{equation}

\no Then, from the AdS-CFT correspondence:

\begin{equation}
\exp(-I_{AdS})\equiv  \left\langle \exp \left(\int d^4 x \,
\left(\,\bar{\chi}^{-}\,\psi_0^-+\bar{\psi}_0^+\,\chi^+\,\right)\right)\right\rangle
\ ,
\end{equation}

\no the unfermion propagator after Fourier transformation to the
coordinate space will be given by \cite{Mueck:1998iz}

\begin{align}
\left \langle \chi^+(x)\, \bar{\chi}^-(y)\right\rangle =
\frac{2\,G}{\pi^2}\; \frac{\Gamma(d_F+\frac12)}{\Gamma
(d_F-\frac32)}\;
\frac{\gamma_\mu(x^\mu-y^\mu)}{|x-y|^{2(d_F+\frac12)}}\,.
\end{align}

On the other hand, the first term in (\ref{I_unfermion}) gives the
contact term interaction $\slash{\partial}\,
\delta(x)$.\footnote{The use of this unfermion contact (or
non-local) interaction is not so clear because there is no fermionic
gauge singlet in the standard model. Gauged unparticle
\cite{Cacciapaglia:2007jq} would be possible, but it is highly
constrained by experiments.} Again, as is the case with the
vector unparticle, the contact term can be removed (at a given
$\epsilon$) by adding the boundary counter term as

\begin{align}
\delta S_{bound} &= \epsilon^{-2(4-d_F)} \int \frac{d^4k}{(2\pi)^4}
\,\left(\,c_1 \,\bar{\psi}^{+}_\epsilon k_\mu \gamma^\mu
\psi^{-}_\epsilon + \cdots \,\right) \cr
 &= \int \frac{d^4k}{(2\pi)^4}\,\,\left(\,c_1 \,\bar{\psi}^{+}_0 k_\mu \gamma^\mu
\psi^{-}_0 + \cdots \,\right) \,,
\end{align}

\no which are localized on the UV-brane.  A natural way to introduce
the cut-off dependence of the boundary counter-term is again
characterized by the naive dimension $\Delta_0$ as $c_1 =
\tilde{C}_1\, \epsilon^{5-2\Delta_0}$. Therefore, the RG evolution
of the contact term is given by

\begin{eqnarray}
C_1(\epsilon) = \tilde{C}_1\,\epsilon^{5-2\Delta_0} \left(\;1-
\left(\frac{\tilde{\epsilon}_1}{\epsilon}\right)^{\gamma}\;\right)
\,, \label{rennn}
\end{eqnarray}

\no where we have introduced the anomalous dimension $\gamma =
2(d_F-\Delta_0)$.

The cut-off dependence \eqref{rennn} is consistent with the
Callan-Symanzik equation for the unfermion interaction. In fact,
$C_1$ will reproduce the solution to the Callan-Symanzik equation by
the identifications $\Delta_0=\frac52$ and $\epsilon \sim
\frac{1}{\mu}$.

As a remark, it is important to note that the non-derivative contact
term $\bar{\psi} \psi$ is {\it not} generated through this
regularization procedure. Thus, the contact interactions
$\bar{O}_{\textrm{SM}} O_{\textrm{SM}}$ and
$\bar{O}_{\textrm{SM}}\,\slash{\partial}\,O_{\textrm{SM}}$ are {\it
not} related at all. This should be contrasted with the vector
unparticle, where the non-derivative contact term is also
introduced.\footnote{Contrary to the remark by
\cite{Grinstein:2008qk}, the interpretation of the identical
coefficient as integrating out the massive field with the propagator
$(\slash{p}+M)/(p^2-M^2)$ does not work here because the
non-derivative contact term vanishes in our approach.}

\section{Conclusions}

In this article, we revisited the unparticle interactions and
propagators from the AdS-CFT point of view. We studied both vector
unparticles and unfermions, revealing the relevant boundary
conditions and RG flows. Our focus is on the contact terms whose
importance was emphasized in \cite{Grinstein:2008qk}, but have been
ignored by previous studies of unparticles. We have shown how the
holographic RG flow can generate such contact terms and their
evolution. This construction is the most natural one in the sense it
is in complete agreement with the field theory discussion
\cite{Grinstein:2008qk}. Our prescription also provides a natural
way to understand the evolution of the boundary local counter terms
under the RG flow.

\section*{Acknowledgements}
The research of Y.~N. is supported in part by NSF grant PHY-0555662
and the UC Berkeley Center for Theoretical Physics. He also thanks
the Yukawa Institute for Theoretical Physics at Kyoto University,
where this work was presented at the YITP-W-08-04 on ``Development
of Quantum Field Theory and String Theory''. C. M. Ho acknowledges
the support from Berkeley Center for Theoretical Physics and the
Croucher Foundation.


\begin{thebibliography}{99}
\bibitem{Georgi:2007ek}
  H.~Georgi,
  Phys.\ Rev.\ Lett.\  {\bf 98}, 221601 (2007)
  [arXiv:hep-ph/0703260];
  H.~Georgi, Phys. Lett. B {\bf 650}, 275 (2007)
  [arXiv:0704.2457 [hep-ph]].

\bibitem{Banks:1981nn}
  T.~Banks and A.~Zaks,
  Nucl.\ Phys.\  B {\bf 196}, 189 (1982).


\bibitem{Mack:1975je}
  G.~Mack,
  Commun.\ Math.\ Phys.\  {\bf 55}, 1 (1977).

\bibitem{Nakayama:2007qu}
  Y.~Nakayama,
  Phys.\ Rev.\  D {\bf 76}, 105009 (2007)
  [arXiv:0707.2451 [hep-ph]].

\bibitem{Grinstein:2008qk}
  B.~Grinstein, K.~Intriligator and I.~Z.~Rothstein,
  Phys. Lett. B {\bf 662}, 367 (2008)
  [arXiv:0801.1140 [hep-ph]].

\bibitem{Stephanov:2007ry}
  M.~A.~Stephanov,
  Phys.\ Rev.\  D {\bf 76}, 035008 (2007)
  [arXiv:0705.3049 [hep-ph]].

\bibitem{Lee:2007xd}
  J.~P.~Lee,
  arXiv:0710.2797 [hep-ph].

\bibitem{Strassler:2008bv}
  M.~J.~Strassler,
  arXiv:0801.0629 [hep-ph].
\bibitem{Cacciapaglia:2008ns}
  G.~Cacciapaglia, G.~Marandella and J.~Terning,
  arXiv:0804.0424 [hep-ph].

\bibitem{Friedland:2009iy}
  A.~Friedland, M.~Giannotti and M.~Graesser,
  arXiv:0902.3676 [hep-th].

\bibitem{Gubser:1998bc}
  S.~S.~Gubser, I.~R.~Klebanov and A.~M.~Polyakov,
  Phys.\ Lett.\  B {\bf 428}, 105 (1998)
  [arXiv:hep-th/9802109].
\bibitem{Witten:1998qj}
  E.~Witten,
  Adv.\ Theor.\ Math.\ Phys.\  {\bf 2}, 253 (1998)
  [arXiv:hep-th/9802150].

\bibitem{Perez} M. Perez-Victoria, JHEP {\bf 0105}, 064 (2001) [arXiv: hep-th/0105048].

\bibitem{Mueck:1998iz}
  W.~Mueck and K.~S.~Viswanathan,
  Phys.\ Rev.\  D {\bf 58}, 106006 (1998)
  [arXiv:hep-th/9805145].
\bibitem{surface-term}
M. Henningson and K. Sfetsos, Phys. Lett. B {\bf 431} 63 (1998)
[arXiv: hep-th/9803251].
\bibitem{Cacciapaglia:2007jq}
  G.~Cacciapaglia, G.~Marandella and J.~Terning,
  JHEP {\bf 0801}, 070 (2008)
  [arXiv:0708.0005 [hep-ph]].

\end{thebibliography}
\end{document}